\documentclass[reprint,amsmath,amssymb,aps,prl,nofootinbib]{revtex4-2}

\usepackage{graphicx}
\usepackage{dcolumn}
\usepackage{bm}

\usepackage[hyperindex,breaklinks]{hyperref}

\usepackage{slashed}
\usepackage{physics}
\usepackage{xcolor}
\usepackage[makeroom]{cancel}

\usepackage[normalem]{ulem}

\newcommand{\dmt}{\Delta m^{\rm kin}_{t\bar{t}}}
\begin{document}

\preprint{APS/123-QED}

\title{Probing CPT invariance with top quarks at the  LHC}

\author{A. Belyaev,$^{1,2}$, L. Cerrito,$^{3,4}$ E. Lunghi,$^5$ S. Moretti,$^{1,2,6}$ and N. Sherrill$^7$}
\affiliation{$^1$School of Physics and Astronomy, 
University of Southampton, Southampton, SO17 1BJ,
United Kingdom}
\affiliation{$^2$Particle Physics Department, Rutherford Appleton Laboratory, Chilton, Didcot, Oxon OX11 0QX, UK}
\affiliation{$^3$INFN Sezione di Roma Tor Vergata, 
Rome 00133, Italy}
\affiliation{$^4$Dipartimento di Fisica, Universit\'a di Roma Tor Vergata, 
Rome 00133, Italy}
\affiliation{$^5$Physics Department, Indiana University,
Bloomington, IN 47405, USA}
\affiliation{$^6$Department of Physics and Astronomy, 
Uppsala University, Box 516, SE-751 20 Uppsala,
Sweden}
\affiliation{$^7$Department of Physics and
Astronomy, University of Sussex, Brighton BN1 9QH, UK}

\begin{abstract}
The first model-independent sensitivity to CPT violation in the top-quark sector is extracted from ATLAS and CMS measurements of the top and antitop kinematical mass difference. 
We find that the temporal component of a CPT-violating background field interacting with the top-quark vector current
is restricted within the interval $[-0.13,0.29]$~GeV at 95\% confidence level.
\end{abstract}

\maketitle

CPT invariance is the symmetry under the simultaneous transformations of charge conjugation (C), parity transformation (P), and time reversal (T). It  is observed to be an exact symmetry of nature at the moment.
According to the well-known CPT theorem~\cite{Bell:1955djs,*Pauli:1955,*Luders:1957bpq}, a
local, unitary, and Lorentz invariant quantum field theory in Minkowski spacetime is CPT invariant.  The  CPT 
symmetry ensures that physical observables, 
including  masses, lifetimes, magnetic moments and cross sections of any particle and its antiparticle are the same.
It should also be stressed that the violation of Lorentz invariance does not necessarily imply the violation of CPT invariance.

Connecting theoretical predictions of CPT invariance violation to particle--antiparticle CPT tests has been carried out for all particle species of the Standard Model (SM) with exception of the top quark.
This work addresses this gap, establishing the first sensitivity to top-sector CPT violation from  top and antitop  mass reconstructions in~ATLAS~and~CMS Collaboration measurements.

Kosteleck\'y and Potting demonstrated that spontaneous CPT violation in string theory may result in remnant observables (likely suppressed by the energy scale of this violation) which can be possibly tested in current experiments~\cite{kp91,*kp95}. The subsequent development of effective field theory (EFT) descriptions~\cite{ck97,*ck98,*cg97,*cg98}, now referred to as the Standard Model Extension (SME),  generated intense interest in testing CPT and Lorentz invariance across a variety of systems~\cite{datatables}. 
Within  the SME framework, CPT violation necessarily implies Lorentz violation, while converse does not necessarily hold.
Lorentz- and CPT-violating coefficients described by the SME can be viewed as generalized tensor background fields.
In view of the discovery of the Higgs boson~\cite{ATLAS:2012yve, CMS:2012qbp} described by a homogeneous and isotropic scalar background with $SU(2)\times U(1)$ quantum numbers, searching for new physics in the form of additional backgrounds is well motivated.

In this study, our focus is on CPT-violating SME operators involving top-quark fields. We suggest using measurements of the kinematically reconstructed top and antitop mass difference, $\Delta m^{\rm kin}_{t\bar{t}}$, 
an observable which is uniquely sensitive to CPT violation, to extract the first constraints on various coefficients responsible for it in the top-quark sector of the SME.  
Dedicated measurements of the kinematically reconstructed mass difference have been performed by D\O~\cite{D0}, CDF~\cite{CDF},  ATLAS~\cite{ATLAS2014}, and CMS~\cite{CMS2012, CMS2017}.\footnote{In Ref.~\cite{Cembranos:2006hj}, earlier D\O~ and CDF top mass measurements~\cite{D0:1997ijd, CDF:1997dxz} were used to place a conservative upper limit on the top and antitop mass difference.} All results are consistent with SM expectations within the experimental uncertainties, and may be translated into constraints on top quark coefficients for CPT violation. We emphasize this approach is conceptually different from tests of Lorentz violation in the top sector~\cite{D0:2012rbu,CMS:2024rcv}, which are insensitive to (leading-order) signatures of CPT violation~\cite{Berger:2015yha}.

In our study, we aim to test gauge invariant and renormalizable CPT-violating SME operators of the form~\cite{ck97, Berger:2015yha}
\begin{align}
\label{SU2basis}
\mathcal{L}^{\text{CPT}-} = &-(a_Q)_{\mu AB}\bar Q_A \gamma^\mu Q_B 
- (a_U)_{\mu AB} \bar U_A \gamma^\mu U_B \nonumber\\
&- (a_D)_{\mu AB} \bar D_A \gamma^\mu D_B.
\end{align}
The odd number of operator Lorentz indices implies a change of sign under a CPT transformation.
Therefore, $\mathcal{L}^{\text{CPT}-}$ is odd
under CPT, which is reflected in its superscript.
We are interested in probing operators involving
the third generation ($A=B=3$), where
\begin{equation}
\label{fields}
Q_3 = \begin{pmatrix} t \\ b \end{pmatrix}_L, \quad U_3 = t_R, \quad D_3 = b_R.
\end{equation}
The relevant coefficients for CPT-violating operators are therefore $(a_{Q})_{\mu 33}, 
(a_{U})_{\mu 33}$, and $(a_{D})_{\mu 33}$. 
{Note that the Lagrangian mass
parameter $m_t$ remains identical for top and antitop quarks  in the presence of CPT violation, in accordance with 
Greenberg's theorem~\cite{Greenberg}.}

A reduction of the number of independent coefficients in Eq.~(\ref{SU2basis}) 
is possible under suitable approximations. In the limit of a zero bottom quark mass $m_b$ (in comparison with $m_t$), the phases of $D_3$ and $Q_3$ fields can 
be independently changed. Thus, the
position-dependent field redefinition 
$D_3 \rightarrow \exp[-i(a_D)_{\mu 33}x^\mu]D_3$ 
can be used to remove the last term in Eq.~\eqref{SU2basis}. 
An analogous redefinition 
of $Q_3$ and $U_3$ with the same phase  $\exp[-i(a_Q)_{\mu 33}x^\mu]$ allows to eliminate the term with $Q_3$ and
shifts the coefficient $(a_{Q})_{\mu 33}$
into the $U_3$ term, 
such that $(a_{U})_{\mu 33}  \rightarrow (a_{U})_{\mu 33} - (a_{Q})_{\mu 33} $. This transformation preserves  the structure of all SM terms and yields an equivalent Lagrangian density expressed in the mass eigenstate basis: 
\begin{equation}
\label{Ltop}
\mathcal{L}_{\rm top}^{\text{CPT}-} =  b_\mu\bar t_R\gamma^\mu t_R,
\end{equation}
where $b_\mu \equiv [(a_Q)_{\mu 33}-(a_U)_{\mu 33}]$ and 
only $t_R$ fields appear. 
Under these field redefinitions, CPT violation is isolated to the top-quark sector and quantified through the differential propagation of  $t_R$ relative to $t_L$.~\footnote{One could have chosen $\exp[-i(a_U)_{\mu 33}x^\mu]$ as phase shift for both $Q_3$ and $U_3$
which would lead to 
\begin{equation}
\mathcal{L'}_{\rm top}^{\text{CPT}-} =  b'_\mu\bar t_L\gamma^\mu t_L,
\end{equation}
with $b'_\mu \equiv [(a_U)_{\mu 33}-(a_Q)_{\mu 33}]= -b_\mu$ and 
only $t_L$ fields involved. For this equivalent formulation the experimental limits on $b_\mu$ will be identical to those for Eq.~(\ref{Ltop}).}

Performing the variational 
procedure including the conventional top kinetic 
terms results in a modified Dirac equation 
\begin{equation}
\label{Diract}
\left[i\slashed{\partial} + \tfrac{1}{2}(1-\gamma_5)\slashed{b} - m_t\right]t = 0.
\end{equation}
Plane-wave solutions imply a quartic 
equation in $p^\mu = (E_t,\vec{p})$
with four distinct solutions linear in $b_\mu$:
\begin{align}
\label{p2}
p^2 = \begin{cases} m_t^2 - p\cdot b \pm [(p\cdot b)^2-m_t^2b^2]^{1/2}\hspace{2mm}\rm{(top)}\\
m_t^2 + p\cdot b \pm [(p\cdot b)^2-m_t^2b^2]^{1/2}\hspace{2mm} \rm{(antitop)} 
\end{cases}
\end{align}
neglecting higher-order terms in $b_\mu$. 
The first and second unconventional terms in each row are associated
with the vector and pseudovector pieces of Eq.~\eqref{Ltop}, respectively, and 
the $\pm$ signs denote states of opposite helicities.
The difference between particle and antiparticle
solutions is obtained by $b_\mu \rightarrow -b_\mu$, 
reflecting the CPT-odd property of Eq.~\eqref{Ltop} 
and the effect of CPT conjugation 
on the plane-wave solutions. Note that potential CPT-violating corrections to top and antitop decay widths are suppressed relative to free-propagation effects~\eqref{p2} by the square of the weak coupling constant and are neglected.

The presence of $b_\mu$ implies
both charge- and helicity-dependent energy eigenvalues.
Top $p$ and antitop $\bar p$ momenta and kinematical masses 
$m_t^{\rm{kin}} \equiv \sqrt{p^2}, m_{\bar t}^{\rm{kin}} \equiv \sqrt{\bar p^2}$ are 
reconstructed through the charge and four-momentum
conservation of final-state decay products.
In the conventional CPT-invariant case, 
$m_t^{\rm{kin}} = m_{\bar t}^{\rm{kin}} = m_t$.
The kinematical mass difference 
\begin{equation}
\label{deltam}
\dmt(p,\lambda_p, \bar{p},\lambda_{\bar p}, m_t, b) 
\equiv m_t^{\rm{kin}} - m_{\bar t}^{\rm{kin}}
\end{equation}
parametrizes a CPT-violating asymmetry, where $\lambda_p$ ($\lambda_{\bar p}$) are the top (antitop) helicities.
In principle, measurements of $\dmt$
could be used to extract $b_\mu$. However, 
this is generically non-trivial because $\dmt$ is time dependent. Implicit time dependence 
enters via event-by-event 
reconstructions of the top and antitop four-momenta.
Explicit time dependence enters through $b_\mu$ directly since the relevant experiments are performed in non-inertial Earth-based laboratories. As a result, the coefficient $b_\mu$ is modulated as a function of the laboratory velocity
and rotation rate~\cite{k98mesons}. It is convenient and standard practice 
to introduce the approximately inertial 
sun-centered frame (SCF) 
where the coefficients for CPT violation carry indices $\mu = \{T, X, Y, Z\}$ and 
may be approximated as constants~\cite{Bluhm:2001rw,*Bluhm:2003un,*Kostelecky:2003fs,*Kostelecky:2002hh}. 
In this setting, the leading laboratory signatures are given by single harmonics of the Earth's sidereal rotation frequency
$\omega_{\oplus} \approx 2\pi/(23\;\rm{h}\;56\;{min})$.
Despite these complications, we demonstrate that 
suitably averaged observables 
permit analyses of the set  $\{b_T, b_X, b_Y, b_Z\}$.

{The ATLAS and CMS Collaborations have reported measurements of  
$\langle \dmt \rangle$~\cite{ATLAS2014,CMS2012,CMS2017}, where $\langle  \rangle$ indicates averaging over all events.} ATLAS used a sample of $t\bar t$ events in the single charged 
lepton + jets decay mode, selected from 4.7 fb$^{-1}$ of $pp$ collisions at $\sqrt{s} = 7$\;TeV~\cite{ATLAS2014}. The data were regularly collected over several months in 2011~\cite{ATLAS2014data}. 
The kinematical mass difference $\langle \dmt \rangle$ was determined
from a maximum likelihood fit to the per-event top and antitop candidate mass difference, reconstructed
in the ATLAS detector frame.
 The reported result
 \begin{align}
 \langle \dmt \rangle^{\rm ATLAS} = 0.67\pm 0.61_{\rm stat}\pm 0.41_{\rm syst}\; {\rm GeV}
 \end{align}
is consistent
with zero within uncertainties. The measurement involved averaging over several sidereal
days and sampling of the full $t\bar t$ phase space. This measurement therefore has negligible sensitivity
to the top polarizations and spatial components of $b_\mu$. The invariance
of the temporal component $b_0$ under the rotation connecting the ATLAS detector
and SCF frames implies $b_0 = b_T$ and Eq.~\eqref{deltam}
takes the form 
\begin{equation}
\langle\dmt\rangle^{\rm{ATLAS}} \approx -\frac{b_T}{m_t}
\frac{\langle E_t + E_{\bar t}\rangle}{2},
\end{equation}
where $\langle E_t + E_{\bar t}\rangle$ is the average over the phase space of the sum of top and antitop energies
characteristic to the dataset. To leading order
in $b_\mu$ the energies $E_t, E_{\bar t}$ are the conventional eigenenergies.

The CMS analyses also selected events where one 
$W$ boson decays hadronically and the other
leptonically. The data were split into 
$\ell^+$ and $\ell^-$ samples. Using the ideogram likelihood method~\cite{ideogram}, 
$m_t^{\rm kin}$ and $m_{\bar t}^{\rm kin}$ were 
reconstructed from the two samples, 
respectively, in the CMS collider frame, from which their difference was obtained. 
The 2017 analysis used a data sample~\cite{CMS2017data} corresponding to $pp$ collisions at $\sqrt{s}=8~$TeV and an integrated luminosity of $19.6\pm 0.5$\;fb$^{-1}$. The data were collected over several months in 2012, yielding~\cite{CMS:2016les}
\begin{align}
\langle \dmt \rangle^{\rm CMS} = -0.15\pm0.19_{\rm stat}\pm 0.09_{\rm syst} \;{\rm GeV}.
\end{align}
Since the tops and antitops were 
selected from different events, CMS is sensitive to
a sum of the averages, \\
$\langle E_t\rangle$ + $\langle E_{\bar t}\rangle$, and therefore 
\begin{equation}
\langle\dmt\rangle^{\rm{CMS}} \approx 
-\frac{b_T}{m_t}\frac{\langle E_t \rangle + \langle E_{\bar t} \rangle}{2}.
\end{equation}
\newcolumntype{P}[1]{>{\centering\arraybackslash}p{#1}}
\newcolumntype{L}[1]{>{\arraybackslash}p{#1}}
\begin{table}[htb]
\centering
\begin{tabular}{|p{2.8cm}| P{1.8cm}| P{1.8cm}| P{1.8cm}|}
\hline
                 &$t\bar{t}$ 
                        &$t\bar{t} \to \ell \nu jj b\bar{b}$	tot
                        &$t\bar{t} \to \ell \nu jj b\bar{b}$ fid
                        \\
\hline
$\langle E_t+E_{\bar{t}}  \rangle_{@7~\text{TeV}}$   & 706.3   &	 708.9    & 658.4\\
$\langle E_t+E_{\bar{t}} \rangle_{@8~\text{TeV}}$    & 738.9   &	742.2     & 674.4\\
$\langle E_t+E_{\bar{t}} \rangle_{@13~\text{TeV}}$   & 878.8   &	883.7     &725.2 \\
$\langle E_t+E_{\bar{t}} \rangle_{@13.6~\text{TeV}}$ & 892.5   &	 898.7    & 729.1\\
\hline
\end{tabular}
\caption{The values of $\langle E_t+E_{\bar{t}} \rangle$ in GeV at  various LHC energies
    for $pp\to t\bar{t}$ processes:
    a) $pp\to t\bar{t}$ ($t\bar{t}$);
    b) $pp\to t\bar{t} \to \ell \nu jj b\bar{b}$
    with no cuts applied (tot);
    c)  $pp\to t\bar{t} \to \ell \nu jj b\bar{b}$
    with fiducial  CMS cuts applied (fid).
     See further details in the text. The events here are generated with CalcHEP.
} \label{tab:Etop}
	\end{table}

In order to extract an upper bound on $b_T$, we need to calculate the average $\langle E_t + E_{\bar t}\rangle$ for $t\bar t$ events in the fiducial region considered by the two experiments. For the same set of cuts, however, there will be no difference between the 
sum of the averages $\langle E_t\rangle$ + $\langle E_{\bar t}\rangle$ and  average of the sum
$\langle E_t + E_{\bar t}\rangle$.
The evaluation  of $\langle  E_t + E_{\bar t} \rangle $
for $t\bar t$ events at various center-of-mass energies is performed with the aid of the Monte Carlo (MC) generator CalcHEP~\cite{Belyaev:2012qa} and cross-checked using Madgraph~\cite{Alwall:2011uj} (interfaced with Pythia8~\cite{Sjostrand:2014zea} and the detector simulator Delphes~\cite{deFavereau:2013fsa}). The results we find are summarized in Table~\ref{tab:Etop}. 
\\
The value of  $b_T$ including uncertainty
reads 
\begin{align}
b_T = & -\frac{2 \langle\dmt\rangle m_t}{\langle E_t + E_{\bar t}\rangle}\nonumber\\
 & \times \left[ 1 \pm 
     \sqrt{\left(\frac{\delta \langle\dmt\rangle}{\langle\dmt\rangle}\right)^2
          +\left(\frac{\delta \langle E_t + E_{\bar t} \rangle }{\langle E_t + E_{\bar t} \rangle} \right)^2} \right].
          \label{eq:bt}
\end{align}
We estimate the uncertainty $ {\delta \langle E_t+E_{\bar{t}} \rangle }$ 
on the predicted   value $\langle E_t+E_{\bar{t}}  \rangle$
from our MC simulations
 by varying the factorization scale within $m_t/2$ and $2 m_t$, as well as using different PDF sets. We find it to be below 5\% for each uncertainty. We have also evaluated the effect of
 the top-quark width on the 
 $\langle E_t+E_{\bar{t}} \rangle$ value 
 by comparing the results for
   $2\to 2$ versus  $2\to 6$ processes at parton level
 (the second versus the third column in Table~\ref{tab:Etop}) and find an increase of only about $+0.5\%$ on
 $\langle E_t+E_{\bar{t}} \rangle$  when the width is taken into account,
which is simply related to a kinematical effect.
 A more detailed analysis of these uncertainties is not relevant for this study since the uncertainty on $b_T$ is completely dominated by the experimental uncertainty on  $\delta \langle \Delta{m_{t\bar t}^{\rm kin}}\rangle$, which is of  order  100\%. Therefore, even a very conservative value of $ {\delta \langle E_t+E_{\bar{t}}  \rangle }$ of  order  10\% will affect the 
 overall uncertainty for $b_T$ determination at the level of only 1\%, as one can see from Eq.~(\ref{eq:bt}). 
 
Combining experimental statistical and systematic uncertainties in quadrature  and using
 Eq.~(\ref{eq:bt}), we find the following 
 exclusion limits on $b_T$:
\begin{align}
\label{bTlimit}
b_T 
\in
\begin{cases}
[-1.10,0.41]\; \text{GeV} & \text{ATLAS @ 7 TeV} \\
[-0.13,0.29]\; \text{GeV} & \text{CMS @ 8 TeV} \\
\end{cases}
\; .
\end{align}
Outside of these intervals  $b_T$ is excluded at 95\%
confidence level.

 \begin{table}[htb]
    \centering
    \begin{tabular}{|c|cc|cc|cc|cc|}
    \hline
    & \multicolumn{2}{c}{7 TeV} & \multicolumn{2}{|c|}{8 TeV} & \multicolumn{2}{|c|}{13 TeV} & \multicolumn{2}{c|}{13.6 TeV}  \\ 
    & tot & fid & tot & fid & tot & fid & tot & fid \\ \hline
$\langle| \Delta_Z|\rangle$ & 72 & 68 & 76 & 70 & 97 & 79 & 100 & 80 \\
$\langle| C_X |\rangle$ & 74 & 69 & 79 & 70 & 103 & 84 & 103 & 81 \\
$\langle|S_X |\rangle$ & 418 & 329 & 451 & 361 & 590 & 416 & 603 & 405 \\
    \hline    
    \end{tabular}
    \caption{Phase space averages (both in the whole phase space and in the CMS fiducial region) for the quantities described in the text in units of GeV. The events here are generated with MadGraph and  showered/hadronized with Pythia8. Finally, Delphes is used to simulate detector effects.}
    \label{tab:MCDelphes}
\end{table}

The spatial components $b_{X,Y,Z}$ can be constrained with the same dataset but require dedicated analyses. 
To good approximation, the laboratory-frame coefficients for both ATLAS and CMS are related to the SCF coefficients via the rotation
\begin{widetext}
\begin{equation}
\label{transformation}
\begin{pmatrix}
b_1 \\ b_2 \\ b_3 
\end{pmatrix}
= 
\begin{pmatrix}-b_Z\sin\chi\cos\psi + \cos(\omega_\oplus T_\oplus)(b_X\cos\chi\cos\psi
+ b_Y\sin\psi) + \sin(\omega_\oplus T_\oplus)(b_Y\cos\chi\cos\psi - b_X\sin\psi) \\
b_Z\cos\chi + \sin\chi[b_X\cos(\omega_\oplus T_\oplus) + b_Y\sin(\omega_\oplus T_\oplus)] \\
-b_Z\sin\chi\sin\psi + \cos(\omega_\oplus T_\oplus)(b_X\cos\chi\sin\psi
- b_Y\cos\psi) + \sin(\omega_\oplus T_\oplus)(b_Y\cos\chi\sin\psi + b_X\cos\psi)
\end{pmatrix},
\end{equation}
\end{widetext}
{where the colatitude is $\chi = 43.7^{\rm o}$ and the beamline orientation north-of-east is $\psi = -11.3^{\rm o}$. The time $T_\oplus$ is identified with the event time. 
Note that this choice of laboratory frame is identical for ATLAS and CMS: the 2-axis points in the upward vertical direction and the 1-axis points towards (away from) the center of the LHC ring for ATLAS (CMS).}

The kinematical mass difference for a single event in terms of the SCF coefficients is given by\footnote{Note that in Eq.~\eqref{eq:dmz-omega}, we omit the helicity-dependent terms from Eq.~\eqref{p2} as they vanish under phase-space and time averages.
}

\begin{widetext}
\begin{align}
\Delta m_{t\bar t}^{\rm kin}  &= - \frac{1}{2m_t} \left[ b_T \Delta_T +  b_Z  \Delta_Z + \sum_{A=X,Y} b_A \Big( C_A \cos(\omega_\oplus T_\oplus) + S_A \sin(\omega_\oplus T_\oplus ) \Big) \right]\; ,
\label{eq:dmz-omega}\\
\Delta_T &=  E_t + E_{\bar t}\; , \\
\Delta_Z &= -\sin\chi\cos\psi(p_1+\bar p_1) + \cos\chi(p_2+\bar p_2) - \sin\chi \sin\psi(p_3 + \bar p_3) \; ,\\
C_X &= \cos\chi\cos\psi(p_1+\bar p_1) + \sin\chi (p_2+\bar p_2)+ \cos\chi\sin\psi(p_3+\bar p_3) \; ,\\
S_X &= -\sin\psi(p_1+\bar p_1) + \cos\psi (p_3+\bar p_3)\;, \\
C_Y &=  - S_X \; ,\\
S_Y &= C_X \; , 
\end{align}
\end{widetext}
where $p_i$ ($\bar{p}_i$) are the top (antitop) three-momentum components in the laboratory frame.
The symmetry of the $pp$ collisions guarantees $\langle p_i + \bar p_i \rangle = 0$, and hence $\langle \Delta m_{t\bar t}^{\rm kin} \rangle = -\frac{b_T}{2m_t} \langle E_t + E_{\bar t} \rangle$ when averaging over the whole phase space. This means that the coefficients $b_{X,Y,Z}$ can only be constrained by dedicated analyses. Moreover, note that the coefficients $b_X$ and $b_Y$ induce effects that vanish also when averaged over time.

A strategy to build an observable sensitive exclusively to $b_Z$ is to consider
\begin{align}
\langle {\Delta m_{t\bar t}^{\rm kin}}' \rangle = 
\langle \Delta m_{t\bar t}^{\rm kin} \; {\rm sgn} [\Delta_Z] \rangle = 
-\frac{b_Z}{2 m_t} \langle | \Delta_Z | \rangle.
\end{align}
In fact, $\langle \Delta_T \; {\rm sgn} [\Delta _Z] \rangle = 0$ and time averaging eliminates $b_{X,Y}$. This quantity can be measured by first calculating $\Delta m_{t\bar t}^{\rm kin} \; {\rm sgn} [\Delta_Z]$ on an event-by-event basis ($\Delta _Z$ is a simple function of the top and antitop three-momenta) and then studying its distribution. Since it is not possible to simply convert the measured $\langle {\Delta m_{t\bar t}^{\rm kin}} \rangle$ distribution into the corresponding $\langle {\Delta m_{t\bar t}^{\rm kin}}' \rangle$ one, a constraint on $b_Z$ can only be obtained via a dedicated re-analysis of the data. The $b_Z$ sensitivity of this analysis is straightforward to assess because, in absence of a signal, we expect the experimental $\langle {\Delta m_{t\bar t}^{\rm kin}} \rangle$ and $\langle {\Delta m_{t\bar t}^{\rm kin}}' \rangle$ distributions to be very similar. In order to get an upper limit on $b_Z$, one should consider the statistical uncertainty of the 8 TeV CMS analysis but adopt the larger ATLAS systematic uncertainty, which has been calculated from a scheme where the top and antitop masses are determined simultaneously on a per-event basis, and center the distribution on zero: $\langle {\Delta m_{t\bar t}^{\rm kin}}' \rangle \sim [0 \pm 0.19 \pm 0.41]\; {\rm GeV}$. Combining these  sensitivities with the MC results for $\langle |\Delta_Z |\rangle $ presented in Table~\ref{tab:MCDelphes} we obtain the expected sensitivity
\begin{align}
\label{bZlimit}
|b_Z|_{\rm expected} \lesssim  4.6\; {\rm GeV}.
\end{align}

For the transverse components, $b_X$ and $b_Y$, the situation is more complicated because a sidereal-time analysis is required. A possible strategy is to divide the sidereal period in a number $N$ of bins. Focusing on the coefficient $b_X$, the kinematical mass difference for events in the $n$-th bin is given by:
\begin{align}
\Delta m_{t\bar t}^{\rm kin}  &= - \frac{b_X}{2m_t} \Delta_X^{(n)}, \\
\Delta_X^{(n)} & = C_X  \langle \cos(\omega_\oplus T_\oplus)\rangle_n + S_X \langle \sin(\omega_\oplus T_\oplus )\rangle_n, 
\end{align}
where $\langle \rangle_n$ indicates the time average over bin $n$. For each bin we can then proceed exactly as for $b_Z$ and average over all events in the bin weighting $\Delta m_{t\bar t}^{\rm kin}$ by ${\rm sgn} [\Delta_X^{(n)}]$. In order to obtain  the sensitivity to the coefficients $b_{X}$ and $b_{Y}$ we simply calculate the phase space average $[\langle |C_A| \rangle   + \langle |S_A| \rangle ]/\sqrt{2}$, where $A=X,Y$, and the factor $1/\sqrt{2}$ is the root-mean-square of the $\sin$ and $\cos$ functions. Following the same procedure as for the $b_Z$ case (but using the appropriate averages from Table~\ref{tab:MCDelphes}), we find the following expected of the sensitivity to $b_{X,Y}$:
\begin{align}
\label{bXYlimits}
|b_{X,Y}|_{\rm expected}  \lesssim  0.8 \; {\rm GeV}.
\end{align}

To summarize, we have established the  first model-independent sensitivity to CPT violation in the top-quark sector using LHC measurements of the top and antitop kinematical mass difference.
The constraint on $b_T$~\eqref{bTlimit} 
is about two orders of magnitude stronger than those expected from single-top production~\cite{Berger:2015yha}.
We have also suggested  dedicated analyses which would lead
to constraints on $|b_Z|$~\eqref{bZlimit} and $|b_{X,Y}|$~\eqref{bXYlimits}  of the same order.
Note that, although constraints on CPT-violating couplings involving $b$ and light quarks from $B$-meson and related oscillations are stringent, they do not directly constrain the third-generation $b$-quark coupling for CPT violation, which is proportional to $(a_Q + a_D)_{33}$. For instance, in $B$-meson oscillations, the observable is governed by the difference $\Delta a = (a_Q + a_D)_{33} - (a_Q + a_D)_{22;\text{or};11}$ across generations~\cite{Kostelecky:2001ff}, which could vanish if CPT violation is flavor universal.
While existing results provide valuable indirect tests of CPT, our work offers a more direct approach focused on the $t$-quark coupling.

Regarding prospects, One might expect LHCb could offer complementary insights via $t\bar{t}$ asymmetry enhancements in the forward kinematic region. However, the quark-antiquark versus gluon-initiated top pair production is not relevant to the observable and analysis proposed in our work, and LHCb cannot fully reconstruct the top-antitop system for the mass difference determination. Therefore, we do not expect complementary insights from LHCb top-quark studies for this specific study. Let us note that an analysis of the the entire Run-2 dataset (140 fb${}^{-1}$), assuming a factor of two improvement on the systematic uncertainty and taking into account the larger $\sqrt{s} = 13\; {\rm TeV}$ center-of-mass energy, is expected to yield a sensitivity to $b_T$ at the 0.05 GeV level (with similar fractional improvements for $b_{X,Y,Z}$). Further accumulation of data will lead to a systematically dominated total uncertainty.

\section*{Acknowledgments}
AB and SM are supported in part through the
NExT Institute and  STFC Consolidated Grant No. ST/L000296/1. SM is also supported by the 
Knut and Alice Wallenberg foundation under the
grant KAW 2017.0100. 
AB acknowledges support from the Leverhulme Trust project MONDMag (RPG-2022-57).
NS is supported in part by the Science and Technology Facilities Council (grant numbers ST/T006048/1 and ST/Y004418/1). EL acknowledges support from the Indiana University Center for Spacetime Symmetries and CERN.
We thank V. A. Kosteleck\'y for useful comments and discussions.

\bibliographystyle{apsrev4-1}
\bibliography{bib}

\begin{thebibliography}{38}%
\makeatletter
\providecommand \@ifxundefined [1]{%
 \@ifx{#1\undefined}
}%
\providecommand \@ifnum [1]{%
 \ifnum #1\expandafter \@firstoftwo
 \else \expandafter \@secondoftwo
 \fi
}%
\providecommand \@ifx [1]{%
 \ifx #1\expandafter \@firstoftwo
 \else \expandafter \@secondoftwo
 \fi
}%
\providecommand \natexlab [1]{#1}%
\providecommand \enquote  [1]{``#1''}%
\providecommand \bibnamefont  [1]{#1}%
\providecommand \bibfnamefont [1]{#1}%
\providecommand \citenamefont [1]{#1}%
\providecommand \href@noop [0]{\@secondoftwo}%
\providecommand \href [0]{\begingroup \@sanitize@url \@href}%
\providecommand \@href[1]{\@@startlink{#1}\@@href}%
\providecommand \@@href[1]{\endgroup#1\@@endlink}%
\providecommand \@sanitize@url [0]{\catcode `\\12\catcode `\$12\catcode
  `\&12\catcode `\#12\catcode `\^12\catcode `\_12\catcode `\%12\relax}%
\providecommand \@@startlink[1]{}%
\providecommand \@@endlink[0]{}%
\providecommand \url  [0]{\begingroup\@sanitize@url \@url }%
\providecommand \@url [1]{\endgroup\@href {#1}{\urlprefix }}%
\providecommand \urlprefix  [0]{URL }%
\providecommand \Eprint [0]{\href }%
\providecommand \doibase [0]{http://dx.doi.org/}%
\providecommand \selectlanguage [0]{\@gobble}%
\providecommand \bibinfo  [0]{\@secondoftwo}%
\providecommand \bibfield  [0]{\@secondoftwo}%
\providecommand \translation [1]{[#1]}%
\providecommand \BibitemOpen [0]{}%
\providecommand \bibitemStop [0]{}%
\providecommand \bibitemNoStop [0]{.\EOS\space}%
\providecommand \EOS [0]{\spacefactor3000\relax}%
\providecommand \BibitemShut  [1]{\csname bibitem#1\endcsname}%
\let\auto@bib@innerbib\@empty
\bibitem [{\citenamefont {Bell}(1955)}]{Bell:1955djs}%
  \BibitemOpen
  \bibfield  {author} {\bibinfo {author} {\bibfnamefont {J.~S.}\ \bibnamefont
  {Bell}},\ }\href {\doibase 10.1098/rspa.1955.0189} {\bibfield  {journal}
  {\bibinfo  {journal} {Proc. Roy. Soc. Lond. A}\ }\textbf {\bibinfo {volume}
  {231}},\ \bibinfo {pages} {479} (\bibinfo {year} {1955})}\BibitemShut
  {NoStop}%
\bibitem [{\citenamefont {Pauli}(1955)}]{Pauli:1955}%
  \BibitemOpen
  \bibfield  {author} {\bibinfo {author} {\bibfnamefont {W.}~\bibnamefont
  {Pauli}},\ }in\ \href@noop {} {\emph {\bibinfo {booktitle} {Niels Bohr and
  the Development of Physics}}},\ \bibinfo {editor} {edited by\ \bibinfo
  {editor} {\bibfnamefont {W.}~\bibnamefont {Pauli}}}\ (\bibinfo  {publisher}
  {McGraw-Hill},\ \bibinfo {address} {New York},\ \bibinfo {year} {1955})\
  p.~\bibinfo {pages} {30}\BibitemShut {NoStop}%
\bibitem [{\citenamefont {Luders}(1957)}]{Luders:1957bpq}%
  \BibitemOpen
  \bibfield  {author} {\bibinfo {author} {\bibfnamefont {G.}~\bibnamefont
  {Luders}},\ }\href {\doibase 10.1016/0003-4916(57)90032-5} {\bibfield
  {journal} {\bibinfo  {journal} {Annals Phys.}\ }\textbf {\bibinfo {volume}
  {2}},\ \bibinfo {pages} {1} (\bibinfo {year} {1957})}\BibitemShut {NoStop}%
\bibitem [{\citenamefont {Kosteleck\'y}\ and\ \citenamefont
  {Potting}(1991)}]{kp91}%
  \BibitemOpen
  \bibfield  {author} {\bibinfo {author} {\bibfnamefont {V.~A.}\ \bibnamefont
  {Kosteleck\'y}}\ and\ \bibinfo {author} {\bibfnamefont {R.}~\bibnamefont
  {Potting}},\ }\href {\doibase 10.1016/0550-3213(91)90071-5} {\bibfield
  {journal} {\bibinfo  {journal} {Nucl. Phys. B}\ }\textbf {\bibinfo {volume}
  {359}},\ \bibinfo {pages} {545} (\bibinfo {year} {1991})}\BibitemShut
  {NoStop}%
\bibitem [{\citenamefont {Kosteleck\'y}\ and\ \citenamefont
  {Potting}(1995)}]{kp95}%
  \BibitemOpen
  \bibfield  {author} {\bibinfo {author} {\bibfnamefont {V.~A.}\ \bibnamefont
  {Kosteleck\'y}}\ and\ \bibinfo {author} {\bibfnamefont {R.}~\bibnamefont
  {Potting}},\ }\href {\doibase 10.1103/PhysRevD.51.3923} {\bibfield  {journal}
  {\bibinfo  {journal} {Phys. Rev. D}\ }\textbf {\bibinfo {volume} {51}},\
  \bibinfo {pages} {3923} (\bibinfo {year} {1995})},\ \Eprint
  {http://arxiv.org/abs/hep-ph/9501341} {arXiv:hep-ph/9501341} \BibitemShut
  {NoStop}%
\bibitem [{\citenamefont {Colladay}\ and\ \citenamefont
  {Kosteleck\'y}(1997)}]{ck97}%
  \BibitemOpen
  \bibfield  {author} {\bibinfo {author} {\bibfnamefont {D.}~\bibnamefont
  {Colladay}}\ and\ \bibinfo {author} {\bibfnamefont {V.~A.}\ \bibnamefont
  {Kosteleck\'y}},\ }\href {\doibase 10.1103/PhysRevD.55.6760} {\bibfield
  {journal} {\bibinfo  {journal} {Phys. Rev. D}\ }\textbf {\bibinfo {volume}
  {55}},\ \bibinfo {pages} {6760} (\bibinfo {year} {1997})},\ \Eprint
  {http://arxiv.org/abs/hep-ph/9703464} {arXiv:hep-ph/9703464} \BibitemShut
  {NoStop}%
\bibitem [{\citenamefont {Colladay}\ and\ \citenamefont
  {Kosteleck\'y}(1998)}]{ck98}%
  \BibitemOpen
  \bibfield  {author} {\bibinfo {author} {\bibfnamefont {D.}~\bibnamefont
  {Colladay}}\ and\ \bibinfo {author} {\bibfnamefont {V.~A.}\ \bibnamefont
  {Kosteleck\'y}},\ }\href {\doibase 10.1103/PhysRevD.58.116002} {\bibfield
  {journal} {\bibinfo  {journal} {Phys. Rev. D}\ }\textbf {\bibinfo {volume}
  {58}},\ \bibinfo {pages} {116002} (\bibinfo {year} {1998})},\ \Eprint
  {http://arxiv.org/abs/hep-ph/9809521} {arXiv:hep-ph/9809521} \BibitemShut
  {NoStop}%
\bibitem [{\citenamefont {Coleman}\ and\ \citenamefont {Glashow}(1997)}]{cg97}%
  \BibitemOpen
  \bibfield  {author} {\bibinfo {author} {\bibfnamefont {S.~R.}\ \bibnamefont
  {Coleman}}\ and\ \bibinfo {author} {\bibfnamefont {S.~L.}\ \bibnamefont
  {Glashow}},\ }\href {\doibase 10.1016/S0370-2693(97)00638-2} {\bibfield
  {journal} {\bibinfo  {journal} {Phys. Lett. B}\ }\textbf {\bibinfo {volume}
  {405}},\ \bibinfo {pages} {249} (\bibinfo {year} {1997})},\ \Eprint
  {http://arxiv.org/abs/hep-ph/9703240} {arXiv:hep-ph/9703240} \BibitemShut
  {NoStop}%
\bibitem [{\citenamefont {Coleman}\ and\ \citenamefont {Glashow}(1999)}]{cg98}%
  \BibitemOpen
  \bibfield  {author} {\bibinfo {author} {\bibfnamefont {S.~R.}\ \bibnamefont
  {Coleman}}\ and\ \bibinfo {author} {\bibfnamefont {S.~L.}\ \bibnamefont
  {Glashow}},\ }\href {\doibase 10.1103/PhysRevD.59.116008} {\bibfield
  {journal} {\bibinfo  {journal} {Phys. Rev. D}\ }\textbf {\bibinfo {volume}
  {59}},\ \bibinfo {pages} {116008} (\bibinfo {year} {1999})},\ \Eprint
  {http://arxiv.org/abs/hep-ph/9812418} {arXiv:hep-ph/9812418} \BibitemShut
  {NoStop}%
\bibitem [{\citenamefont {Kosteleck\'y}\ and\ \citenamefont
  {Russell}(2011)}]{datatables}%
  \BibitemOpen
  \bibfield  {author} {\bibinfo {author} {\bibfnamefont {V.~A.}\ \bibnamefont
  {Kosteleck\'y}}\ and\ \bibinfo {author} {\bibfnamefont {N.}~\bibnamefont
  {Russell}},\ }\href {\doibase 10.1103/RevModPhys.83.11} {\bibfield  {journal}
  {\bibinfo  {journal} {Rev. Mod. Phys.}\ }\textbf {\bibinfo {volume} {83}},\
  \bibinfo {pages} {11} (\bibinfo {year} {2011})},\ \Eprint
  {http://arxiv.org/abs/0801.0287} {arXiv:0801.0287 [hep-ph]} \BibitemShut
  {NoStop}%
\bibitem [{\citenamefont {Aad}\ \emph {et~al.}(2012)\citenamefont {Aad} \emph
  {et~al.}}]{ATLAS:2012yve}%
  \BibitemOpen
  \bibfield  {author} {\bibinfo {author} {\bibfnamefont {G.}~\bibnamefont
  {Aad}} \emph {et~al.} (\bibinfo {collaboration} {ATLAS}),\ }\href {\doibase
  10.1016/j.physletb.2012.08.020} {\bibfield  {journal} {\bibinfo  {journal}
  {Phys. Lett. B}\ }\textbf {\bibinfo {volume} {716}},\ \bibinfo {pages} {1}
  (\bibinfo {year} {2012})},\ \Eprint {http://arxiv.org/abs/1207.7214}
  {arXiv:1207.7214 [hep-ex]} \BibitemShut {NoStop}%
\bibitem [{\citenamefont {Chatrchyan}\ \emph
  {et~al.}(2012{\natexlab{a}})\citenamefont {Chatrchyan} \emph
  {et~al.}}]{CMS:2012qbp}%
  \BibitemOpen
  \bibfield  {author} {\bibinfo {author} {\bibfnamefont {S.}~\bibnamefont
  {Chatrchyan}} \emph {et~al.} (\bibinfo {collaboration} {CMS}),\ }\href
  {\doibase 10.1016/j.physletb.2012.08.021} {\bibfield  {journal} {\bibinfo
  {journal} {Phys. Lett. B}\ }\textbf {\bibinfo {volume} {716}},\ \bibinfo
  {pages} {30} (\bibinfo {year} {2012}{\natexlab{a}})},\ \Eprint
  {http://arxiv.org/abs/1207.7235} {arXiv:1207.7235 [hep-ex]} \BibitemShut
  {NoStop}%
\bibitem [{\citenamefont {Abazov}\ \emph {et~al.}(2011)\citenamefont {Abazov}
  \emph {et~al.}}]{D0}%
  \BibitemOpen
  \bibfield  {author} {\bibinfo {author} {\bibfnamefont {V.~M.}\ \bibnamefont
  {Abazov}} \emph {et~al.} (\bibinfo {collaboration} {D0}),\ }\href {\doibase
  10.1103/PhysRevD.84.052005} {\bibfield  {journal} {\bibinfo  {journal} {Phys.
  Rev. D}\ }\textbf {\bibinfo {volume} {84}},\ \bibinfo {pages} {052005}
  (\bibinfo {year} {2011})},\ \Eprint {http://arxiv.org/abs/1106.2063}
  {arXiv:1106.2063 [hep-ex]} \BibitemShut {NoStop}%
\bibitem [{\citenamefont {Aaltonen}\ \emph {et~al.}(2013)\citenamefont
  {Aaltonen} \emph {et~al.}}]{CDF}%
  \BibitemOpen
  \bibfield  {author} {\bibinfo {author} {\bibfnamefont {T.}~\bibnamefont
  {Aaltonen}} \emph {et~al.} (\bibinfo {collaboration} {CDF}),\ }\href
  {\doibase 10.1103/PhysRevD.87.052013} {\bibfield  {journal} {\bibinfo
  {journal} {Phys. Rev. D}\ }\textbf {\bibinfo {volume} {87}},\ \bibinfo
  {pages} {052013} (\bibinfo {year} {2013})},\ \Eprint
  {http://arxiv.org/abs/1210.6131} {arXiv:1210.6131 [hep-ex]} \BibitemShut
  {NoStop}%
\bibitem [{\citenamefont {Aad}\ \emph {et~al.}(2014)\citenamefont {Aad} \emph
  {et~al.}}]{ATLAS2014}%
  \BibitemOpen
  \bibfield  {author} {\bibinfo {author} {\bibfnamefont {G.}~\bibnamefont
  {Aad}} \emph {et~al.} (\bibinfo {collaboration} {ATLAS}),\ }\href {\doibase
  10.1016/j.physletb.2013.12.010} {\bibfield  {journal} {\bibinfo  {journal}
  {Phys. Lett. B}\ }\textbf {\bibinfo {volume} {728}},\ \bibinfo {pages} {363}
  (\bibinfo {year} {2014})},\ \Eprint {http://arxiv.org/abs/1310.6527}
  {arXiv:1310.6527 [hep-ex]} \BibitemShut {NoStop}%
\bibitem [{\citenamefont {Chatrchyan}\ \emph
  {et~al.}(2012{\natexlab{b}})\citenamefont {Chatrchyan} \emph
  {et~al.}}]{CMS2012}%
  \BibitemOpen
  \bibfield  {author} {\bibinfo {author} {\bibfnamefont {S.}~\bibnamefont
  {Chatrchyan}} \emph {et~al.} (\bibinfo {collaboration} {CMS}),\ }\href
  {\doibase 10.1007/JHEP06(2012)109} {\bibfield  {journal} {\bibinfo  {journal}
  {JHEP}\ }\textbf {\bibinfo {volume} {06}},\ \bibinfo {pages} {109} (\bibinfo
  {year} {2012}{\natexlab{b}})},\ \Eprint {http://arxiv.org/abs/1204.2807}
  {arXiv:1204.2807 [hep-ex]} \BibitemShut {NoStop}%
\bibitem [{\citenamefont {Chatrchyan}\ \emph
  {et~al.}(2017{\natexlab{a}})\citenamefont {Chatrchyan} \emph
  {et~al.}}]{CMS2017}%
  \BibitemOpen
  \bibfield  {author} {\bibinfo {author} {\bibfnamefont {S.}~\bibnamefont
  {Chatrchyan}} \emph {et~al.} (\bibinfo {collaboration} {CMS}),\ }\href
  {\doibase 10.1016/j.physletb.2017.04.028} {\bibfield  {journal} {\bibinfo
  {journal} {Phys. Lett. B}\ }\textbf {\bibinfo {volume} {770}},\ \bibinfo
  {pages} {50} (\bibinfo {year} {2017}{\natexlab{a}})},\ \Eprint
  {http://arxiv.org/abs/1610.09551} {arXiv:1610.09551 [hep-ex]} \BibitemShut
  {NoStop}%
\bibitem [{\citenamefont {Cembranos}\ \emph {et~al.}(2008)\citenamefont
  {Cembranos}, \citenamefont {Rajaraman},\ and\ \citenamefont
  {Takayama}}]{Cembranos:2006hj}%
  \BibitemOpen
  \bibfield  {author} {\bibinfo {author} {\bibfnamefont {J.~A.~R.}\
  \bibnamefont {Cembranos}}, \bibinfo {author} {\bibfnamefont {A.}~\bibnamefont
  {Rajaraman}}, \ and\ \bibinfo {author} {\bibfnamefont {F.}~\bibnamefont
  {Takayama}},\ }\href {\doibase 10.1209/0295-5075/82/21001} {\bibfield
  {journal} {\bibinfo  {journal} {EPL}\ }\textbf {\bibinfo {volume} {82}},\
  \bibinfo {pages} {21001} (\bibinfo {year} {2008})},\ \Eprint
  {http://arxiv.org/abs/hep-ph/0609244} {arXiv:hep-ph/0609244} \BibitemShut
  {NoStop}%
\bibitem [{\citenamefont {Abachi}\ \emph {et~al.}(1997)\citenamefont {Abachi}
  \emph {et~al.}}]{D0:1997ijd}%
  \BibitemOpen
  \bibfield  {author} {\bibinfo {author} {\bibfnamefont {S.}~\bibnamefont
  {Abachi}} \emph {et~al.} (\bibinfo {collaboration} {D0}),\ }\href {\doibase
  10.1103/PhysRevLett.79.1197} {\bibfield  {journal} {\bibinfo  {journal}
  {Phys. Rev. Lett.}\ }\textbf {\bibinfo {volume} {79}},\ \bibinfo {pages}
  {1197} (\bibinfo {year} {1997})},\ \Eprint
  {http://arxiv.org/abs/hep-ex/9703008} {arXiv:hep-ex/9703008} \BibitemShut
  {NoStop}%
\bibitem [{\citenamefont {Abe}\ \emph {et~al.}(1998)\citenamefont {Abe} \emph
  {et~al.}}]{CDF:1997dxz}%
  \BibitemOpen
  \bibfield  {author} {\bibinfo {author} {\bibfnamefont {F.}~\bibnamefont
  {Abe}} \emph {et~al.} (\bibinfo {collaboration} {CDF}),\ }\href {\doibase
  10.1103/PhysRevLett.80.2767} {\bibfield  {journal} {\bibinfo  {journal}
  {Phys. Rev. Lett.}\ }\textbf {\bibinfo {volume} {80}},\ \bibinfo {pages}
  {2767} (\bibinfo {year} {1998})},\ \Eprint
  {http://arxiv.org/abs/hep-ex/9801014} {arXiv:hep-ex/9801014} \BibitemShut
  {NoStop}%
\bibitem [{\citenamefont {Abazov}\ \emph {et~al.}(2012)\citenamefont {Abazov}
  \emph {et~al.}}]{D0:2012rbu}%
  \BibitemOpen
  \bibfield  {author} {\bibinfo {author} {\bibfnamefont {V.~M.}\ \bibnamefont
  {Abazov}} \emph {et~al.} (\bibinfo {collaboration} {D0}),\ }\href {\doibase
  10.1103/PhysRevLett.108.261603} {\bibfield  {journal} {\bibinfo  {journal}
  {Phys. Rev. Lett.}\ }\textbf {\bibinfo {volume} {108}},\ \bibinfo {pages}
  {261603} (\bibinfo {year} {2012})},\ \Eprint {http://arxiv.org/abs/1203.6106}
  {arXiv:1203.6106 [hep-ex]} \BibitemShut {NoStop}%
\bibitem [{\citenamefont {Hayrapetyan}\ \emph {et~al.}(2024)\citenamefont
  {Hayrapetyan} \emph {et~al.}}]{CMS:2024rcv}%
  \BibitemOpen
  \bibfield  {author} {\bibinfo {author} {\bibfnamefont {A.}~\bibnamefont
  {Hayrapetyan}} \emph {et~al.} (\bibinfo {collaboration} {CMS}),\ }\href@noop
  {} {\  (\bibinfo {year} {2024})},\ \Eprint {http://arxiv.org/abs/2405.14757}
  {arXiv:2405.14757 [hep-ex]} \BibitemShut {NoStop}%
\bibitem [{\citenamefont {Berger}\ \emph {et~al.}(2016)\citenamefont {Berger},
  \citenamefont {Kosteleck\'y},\ and\ \citenamefont {Liu}}]{Berger:2015yha}%
  \BibitemOpen
  \bibfield  {author} {\bibinfo {author} {\bibfnamefont {M.~S.}\ \bibnamefont
  {Berger}}, \bibinfo {author} {\bibfnamefont {V.~A.}\ \bibnamefont
  {Kosteleck\'y}}, \ and\ \bibinfo {author} {\bibfnamefont {Z.}~\bibnamefont
  {Liu}},\ }\href {\doibase 10.1103/PhysRevD.93.036005} {\bibfield  {journal}
  {\bibinfo  {journal} {Phys. Rev. D}\ }\textbf {\bibinfo {volume} {93}},\
  \bibinfo {pages} {036005} (\bibinfo {year} {2016})},\ \Eprint
  {http://arxiv.org/abs/1509.08929} {arXiv:1509.08929 [hep-ph]} \BibitemShut
  {NoStop}%
\bibitem [{\citenamefont {Greenberg}(2002)}]{Greenberg}%
  \BibitemOpen
  \bibfield  {author} {\bibinfo {author} {\bibfnamefont {O.~W.}\ \bibnamefont
  {Greenberg}},\ }\href {\doibase 10.1103/PhysRevLett.89.231602} {\bibfield
  {journal} {\bibinfo  {journal} {Phys. Rev. Lett.}\ }\textbf {\bibinfo
  {volume} {89}},\ \bibinfo {pages} {231602} (\bibinfo {year} {2002})},\
  \Eprint {http://arxiv.org/abs/hep-ph/0201258} {arXiv:hep-ph/0201258}
  \BibitemShut {NoStop}%
\bibitem [{\citenamefont {Kosteleck\'y}(1998)}]{k98mesons}%
  \BibitemOpen
  \bibfield  {author} {\bibinfo {author} {\bibfnamefont {V.~A.}\ \bibnamefont
  {Kosteleck\'y}},\ }\href {\doibase 10.1103/PhysRevLett.80.1818} {\bibfield
  {journal} {\bibinfo  {journal} {Phys. Rev. Lett.}\ }\textbf {\bibinfo
  {volume} {80}},\ \bibinfo {pages} {1818} (\bibinfo {year} {1998})},\ \Eprint
  {http://arxiv.org/abs/hep-ph/9809572} {arXiv:hep-ph/9809572} \BibitemShut
  {NoStop}%
\bibitem [{\citenamefont {Bluhm}\ \emph {et~al.}(2002)\citenamefont {Bluhm},
  \citenamefont {Kosteleck\'y}, \citenamefont {Lane},\ and\ \citenamefont
  {Russell}}]{Bluhm:2001rw}%
  \BibitemOpen
  \bibfield  {author} {\bibinfo {author} {\bibfnamefont {R.}~\bibnamefont
  {Bluhm}}, \bibinfo {author} {\bibfnamefont {V.~A.}\ \bibnamefont
  {Kosteleck\'y}}, \bibinfo {author} {\bibfnamefont {C.~D.}\ \bibnamefont
  {Lane}}, \ and\ \bibinfo {author} {\bibfnamefont {N.}~\bibnamefont
  {Russell}},\ }\href {\doibase 10.1103/PhysRevLett.88.090801} {\bibfield
  {journal} {\bibinfo  {journal} {Phys. Rev. Lett.}\ }\textbf {\bibinfo
  {volume} {88}},\ \bibinfo {pages} {090801} (\bibinfo {year} {2002})},\
  \Eprint {http://arxiv.org/abs/hep-ph/0111141} {arXiv:hep-ph/0111141}
  \BibitemShut {NoStop}%
\bibitem [{\citenamefont {Bluhm}\ \emph {et~al.}(2003)\citenamefont {Bluhm},
  \citenamefont {Kosteleck\'y}, \citenamefont {Lane},\ and\ \citenamefont
  {Russell}}]{Bluhm:2003un}%
  \BibitemOpen
  \bibfield  {author} {\bibinfo {author} {\bibfnamefont {R.}~\bibnamefont
  {Bluhm}}, \bibinfo {author} {\bibfnamefont {V.~A.}\ \bibnamefont
  {Kosteleck\'y}}, \bibinfo {author} {\bibfnamefont {C.~D.}\ \bibnamefont
  {Lane}}, \ and\ \bibinfo {author} {\bibfnamefont {N.}~\bibnamefont
  {Russell}},\ }\href {\doibase 10.1103/PhysRevD.68.125008} {\bibfield
  {journal} {\bibinfo  {journal} {Phys. Rev. D}\ }\textbf {\bibinfo {volume}
  {68}},\ \bibinfo {pages} {125008} (\bibinfo {year} {2003})},\ \Eprint
  {http://arxiv.org/abs/hep-ph/0306190} {arXiv:hep-ph/0306190} \BibitemShut
  {NoStop}%
\bibitem [{\citenamefont {Kosteleck\'y}(2004)}]{Kostelecky:2003fs}%
  \BibitemOpen
  \bibfield  {author} {\bibinfo {author} {\bibfnamefont {V.~A.}\ \bibnamefont
  {Kosteleck\'y}},\ }\href {\doibase 10.1103/PhysRevD.69.105009} {\bibfield
  {journal} {\bibinfo  {journal} {Phys. Rev. D}\ }\textbf {\bibinfo {volume}
  {69}},\ \bibinfo {pages} {105009} (\bibinfo {year} {2004})},\ \Eprint
  {http://arxiv.org/abs/hep-th/0312310} {arXiv:hep-th/0312310} \BibitemShut
  {NoStop}%
\bibitem [{\citenamefont {Kosteleck\'y}\ and\ \citenamefont
  {Mewes}(2002)}]{Kostelecky:2002hh}%
  \BibitemOpen
  \bibfield  {author} {\bibinfo {author} {\bibfnamefont {V.~A.}\ \bibnamefont
  {Kosteleck\'y}}\ and\ \bibinfo {author} {\bibfnamefont {M.}~\bibnamefont
  {Mewes}},\ }\href {\doibase 10.1103/PhysRevD.66.056005} {\bibfield  {journal}
  {\bibinfo  {journal} {Phys. Rev. D}\ }\textbf {\bibinfo {volume} {66}},\
  \bibinfo {pages} {056005} (\bibinfo {year} {2002})},\ \Eprint
  {http://arxiv.org/abs/hep-ph/0205211} {arXiv:hep-ph/0205211} \BibitemShut
  {NoStop}%
\bibitem [{\citenamefont {Aad}\ \emph {et~al.}(2013)\citenamefont {Aad} \emph
  {et~al.}}]{ATLAS2014data}%
  \BibitemOpen
  \bibfield  {author} {\bibinfo {author} {\bibfnamefont {G.}~\bibnamefont
  {Aad}} \emph {et~al.} (\bibinfo {collaboration} {ATLAS}),\ }\href {\doibase
  10.1140/epjc/s10052-013-2518-3} {\bibfield  {journal} {\bibinfo  {journal}
  {Eur. Phys. J. C}\ }\textbf {\bibinfo {volume} {73}},\ \bibinfo {pages}
  {2518} (\bibinfo {year} {2013})},\ \Eprint {http://arxiv.org/abs/1302.4393}
  {arXiv:1302.4393 [hep-ex]} \BibitemShut {NoStop}%
\bibitem [{\citenamefont {Abdallah}\ \emph {et~al.}(2008)\citenamefont
  {Abdallah} \emph {et~al.}}]{ideogram}%
  \BibitemOpen
  \bibfield  {author} {\bibinfo {author} {\bibfnamefont {J.}~\bibnamefont
  {Abdallah}} \emph {et~al.} (\bibinfo {collaboration} {DELPHI}),\ }\href
  {\doibase 10.1140/epjc/s10052-008-0585-7} {\bibfield  {journal} {\bibinfo
  {journal} {Eur. Phys. J. C}\ }\textbf {\bibinfo {volume} {55}},\ \bibinfo
  {pages} {1} (\bibinfo {year} {2008})},\ \Eprint
  {http://arxiv.org/abs/0803.2534} {arXiv:0803.2534 [hep-ex]} \BibitemShut
  {NoStop}%
\bibitem [{\citenamefont {Chatrchyan}\ \emph {et~al.}(2013)\citenamefont
  {Chatrchyan} \emph {et~al.}}]{CMS2017data}%
  \BibitemOpen
  \bibfield  {author} {\bibinfo {author} {\bibfnamefont {S.}~\bibnamefont
  {Chatrchyan}} \emph {et~al.} (\bibinfo {collaboration} {CMS}),\ }\href@noop
  {} {\bibfield  {journal} {\bibinfo  {journal} {CMS-PAS-LUM-13-001}\ }
  (\bibinfo {year} {2013})}\BibitemShut {NoStop}%
\bibitem [{\citenamefont {Chatrchyan}\ \emph
  {et~al.}(2017{\natexlab{b}})\citenamefont {Chatrchyan} \emph
  {et~al.}}]{CMS:2016les}%
  \BibitemOpen
  \bibfield  {author} {\bibinfo {author} {\bibfnamefont {S.}~\bibnamefont
  {Chatrchyan}} \emph {et~al.} (\bibinfo {collaboration} {CMS}),\ }\href
  {\doibase 10.1016/j.physletb.2017.04.028} {\bibfield  {journal} {\bibinfo
  {journal} {Phys. Lett. B}\ }\textbf {\bibinfo {volume} {770}},\ \bibinfo
  {pages} {50} (\bibinfo {year} {2017}{\natexlab{b}})},\ \Eprint
  {http://arxiv.org/abs/1610.09551} {arXiv:1610.09551 [hep-ex]} \BibitemShut
  {NoStop}%
\bibitem [{\citenamefont {Belyaev}\ \emph {et~al.}(2013)\citenamefont
  {Belyaev}, \citenamefont {Christensen},\ and\ \citenamefont
  {Pukhov}}]{Belyaev:2012qa}%
  \BibitemOpen
  \bibfield  {author} {\bibinfo {author} {\bibfnamefont {A.}~\bibnamefont
  {Belyaev}}, \bibinfo {author} {\bibfnamefont {N.~D.}\ \bibnamefont
  {Christensen}}, \ and\ \bibinfo {author} {\bibfnamefont {A.}~\bibnamefont
  {Pukhov}},\ }\href {\doibase 10.1016/j.cpc.2013.01.014} {\bibfield  {journal}
  {\bibinfo  {journal} {Comput. Phys. Commun.}\ }\textbf {\bibinfo {volume}
  {184}},\ \bibinfo {pages} {1729} (\bibinfo {year} {2013})},\ \Eprint
  {http://arxiv.org/abs/1207.6082} {arXiv:1207.6082 [hep-ph]} \BibitemShut
  {NoStop}%
\bibitem [{\citenamefont {Alwall}\ \emph {et~al.}(2011)\citenamefont {Alwall},
  \citenamefont {Herquet}, \citenamefont {Maltoni}, \citenamefont {Mattelaer},\
  and\ \citenamefont {Stelzer}}]{Alwall:2011uj}%
  \BibitemOpen
  \bibfield  {author} {\bibinfo {author} {\bibfnamefont {J.}~\bibnamefont
  {Alwall}}, \bibinfo {author} {\bibfnamefont {M.}~\bibnamefont {Herquet}},
  \bibinfo {author} {\bibfnamefont {F.}~\bibnamefont {Maltoni}}, \bibinfo
  {author} {\bibfnamefont {O.}~\bibnamefont {Mattelaer}}, \ and\ \bibinfo
  {author} {\bibfnamefont {T.}~\bibnamefont {Stelzer}},\ }\href {\doibase
  10.1007/JHEP06(2011)128} {\bibfield  {journal} {\bibinfo  {journal} {JHEP}\
  }\textbf {\bibinfo {volume} {06}},\ \bibinfo {pages} {128} (\bibinfo {year}
  {2011})},\ \Eprint {http://arxiv.org/abs/1106.0522} {arXiv:1106.0522
  [hep-ph]} \BibitemShut {NoStop}%
\bibitem [{\citenamefont {Sj\"ostrand}\ \emph {et~al.}(2015)\citenamefont
  {Sj\"ostrand}, \citenamefont {Ask}, \citenamefont {Christiansen},
  \citenamefont {Corke}, \citenamefont {Desai}, \citenamefont {Ilten},
  \citenamefont {Mrenna}, \citenamefont {Prestel}, \citenamefont {Rasmussen},\
  and\ \citenamefont {Skands}}]{Sjostrand:2014zea}%
  \BibitemOpen
  \bibfield  {author} {\bibinfo {author} {\bibfnamefont {T.}~\bibnamefont
  {Sj\"ostrand}}, \bibinfo {author} {\bibfnamefont {S.}~\bibnamefont {Ask}},
  \bibinfo {author} {\bibfnamefont {J.~R.}\ \bibnamefont {Christiansen}},
  \bibinfo {author} {\bibfnamefont {R.}~\bibnamefont {Corke}}, \bibinfo
  {author} {\bibfnamefont {N.}~\bibnamefont {Desai}}, \bibinfo {author}
  {\bibfnamefont {P.}~\bibnamefont {Ilten}}, \bibinfo {author} {\bibfnamefont
  {S.}~\bibnamefont {Mrenna}}, \bibinfo {author} {\bibfnamefont
  {S.}~\bibnamefont {Prestel}}, \bibinfo {author} {\bibfnamefont {C.~O.}\
  \bibnamefont {Rasmussen}}, \ and\ \bibinfo {author} {\bibfnamefont {P.~Z.}\
  \bibnamefont {Skands}},\ }\href {\doibase 10.1016/j.cpc.2015.01.024}
  {\bibfield  {journal} {\bibinfo  {journal} {Comput. Phys. Commun.}\ }\textbf
  {\bibinfo {volume} {191}},\ \bibinfo {pages} {159} (\bibinfo {year}
  {2015})},\ \Eprint {http://arxiv.org/abs/1410.3012} {arXiv:1410.3012
  [hep-ph]} \BibitemShut {NoStop}%
\bibitem [{\citenamefont {de~Favereau}\ \emph {et~al.}(2014)\citenamefont
  {de~Favereau}, \citenamefont {Delaere}, \citenamefont {Demin}, \citenamefont
  {Giammanco}, \citenamefont {Lema\^\i{}tre}, \citenamefont {Mertens},\ and\
  \citenamefont {Selvaggi}}]{deFavereau:2013fsa}%
  \BibitemOpen
  \bibfield  {author} {\bibinfo {author} {\bibfnamefont {J.}~\bibnamefont
  {de~Favereau}}, \bibinfo {author} {\bibfnamefont {C.}~\bibnamefont
  {Delaere}}, \bibinfo {author} {\bibfnamefont {P.}~\bibnamefont {Demin}},
  \bibinfo {author} {\bibfnamefont {A.}~\bibnamefont {Giammanco}}, \bibinfo
  {author} {\bibfnamefont {V.}~\bibnamefont {Lema\^\i{}tre}}, \bibinfo {author}
  {\bibfnamefont {A.}~\bibnamefont {Mertens}}, \ and\ \bibinfo {author}
  {\bibfnamefont {M.}~\bibnamefont {Selvaggi}} (\bibinfo {collaboration}
  {DELPHES 3}),\ }\href {\doibase 10.1007/JHEP02(2014)057} {\bibfield
  {journal} {\bibinfo  {journal} {JHEP}\ }\textbf {\bibinfo {volume} {02}},\
  \bibinfo {pages} {057} (\bibinfo {year} {2014})},\ \Eprint
  {http://arxiv.org/abs/1307.6346} {arXiv:1307.6346 [hep-ex]} \BibitemShut
  {NoStop}%
\bibitem [{\citenamefont {Kostelecky}(2001)}]{Kostelecky:2001ff}%
  \BibitemOpen
  \bibfield  {author} {\bibinfo {author} {\bibfnamefont {V.~A.}\ \bibnamefont
  {Kostelecky}},\ }\href {\doibase 10.1103/PhysRevD.64.076001} {\bibfield
  {journal} {\bibinfo  {journal} {Phys. Rev. D}\ }\textbf {\bibinfo {volume}
  {64}},\ \bibinfo {pages} {076001} (\bibinfo {year} {2001})},\ \Eprint
  {http://arxiv.org/abs/hep-ph/0104120} {arXiv:hep-ph/0104120} \BibitemShut
  {NoStop}%
\end{thebibliography}%

\end{document}